\documentclass[a4paper,11pt]{article}
\usepackage[dvips]{graphicx}
\usepackage{amssymb}
\usepackage{amsmath}

\usepackage{cite}

\begin{document}

\title{An Exponential F(R) Dark Energy Model}
\author{V. K. Oikonomou\thanks{
voiko@physics.auth.gr}\\
Max Planck Institute for Mathematics in the Sciences\\
Inselstrasse 22, 04103 Leipzig, Germany} \maketitle

\begin{abstract}
We present an exponential $F(R)$ modified gravity model in the Jordan and the Einstein frame. We use a general approach in order to investigate and demonstrate the viability of the model. Apart from the general features that this models has, which actually render it viable at a first step, we address the issues of finite time singularities, Newton's law corrections and the scalaron mass. As we will evince, the model passes these latter two tests successfully and also has no finite time singularities, a feature inherent to other well studied exponential models.
\end{abstract}

\section*{Introduction}

General Relativity is one of the most sound scientific
descriptions of nature, regarding gravitational interactions in
scales where gravity becomes important. Hence, it has become a
powerful tool to describe local gravitational interactions and
moreover to describe how the universe evolves as a whole. The
astrophysical observations of the late 90's have put into a
different perspective the way that we think the universe is
evolving. According to these observations, the universe has
undergone two accelerating phases. The first was the inflation
period, while the second is the present epoch's acceleration. In
reference to the latter, this cosmic acceleration is known as dark
energy. The observational data for the present epoch (the new
results of Planck telescope) suggest that the universe is
described by the $\mathrm{\Lambda}\mathrm{CDM}$ model. The main
features of this model is that the universe is almost spatially
flat, and consists of ordinary matter ($\sim4.9\%$), cold dark
matter ($\sim 26.8\%$) and dark energy ($\sim 68.3\%$), with the
latter being the reason behind the present day acceleration.

\noindent One of the most recent attempts to successfully describe
the dark energy is provided by the $F(R)$ modified theories of
gravity (for informative reviews and very important papers on
these theories see \cite{reviews,importantpapers,paper1,paper2}),
in the context of which, what changes drastically is not the left
hand side of the Einstein equations, but the right hand side. In
order the Friedmann-Robertson-Walker equations give an
accelerating solution, the energy momentum tensor must contain a
fluid with negative $w$, and this is achieved with these $F(R)$
theories. Particularly, early time inflation and the late time
acceleration enjoy a unified description within the
self-consistent theoretical framework that some of these models
provide
\cite{reviews,importantpapers,paper1,paper2,generalfrtheories,singularitiesaccelerationmodels,exponentialmodels}.
In addition, some of these models are put into a string theory
oriented context \cite{stringmodifiedgravities} and also quantum
corrections have been calculated \cite{quantumcorrections}.
Connections to other gravitational extensions of general
relativity, such as Gauss-Bonnet gravity, have also been pointed
out in the literature \cite{gaussbonet} and moreover some exact
solutions in strong gravitational background have been done in
\cite{solutions}.

\noindent A very interesting and novel feature that most of the
$F(R)$ theories of gravity have, is the appearance of finite time
singularities in the physical parameters. There are several types
of singularities depending on which physical parameters become
singular. This finite type singularities can completely change our
universe's evolutionary history, leading to rather dramatic final
states, known as Big Rip, or Little Rip etc. For an important
stream of papers in regard to cosmological singularities and
related issues, see \cite{singularitiesaccelerationmodels,reviews}
and references therein.

 With the $F(R)$ theories being a generalization of general relativity, inevitably, these theories are confronted with the successes of general relativity. This entails a number of constraints that need to be satisfied, in order a promising modified gravity theory can be considered viable. These constraints are related to the local tests of general relativity and additionally to various cosmological bounds. The local tests are related to planetary and star formation tests. Moreover, each $F(R)$ theory formulated in the Jordan frame has  a corresponding scalar-tensor gravitational theory in the Einstein frame, the scalarons of which have to be classical, so that the theory is quantum-mechanically stable.

 In this article, we shall analyze in detail one exponential model which has some very appealing features in reference to the constraints we mentioned above. Exponential models have been thoroughly studied in references \cite{paper1,paper2,exponentialmodels}. One of the most appealing features of the exponential models is that, in some of them, singularities are absent, while at the same time these models successfully pass all the tests these models are confronted with. The model we shall present has very interesting attributes, since it is free of singularities. Also we address the matter instability and Newton's law corrections issues and as we will demonstrate, the model at hand passes these tests successfully. However, the model has four free parameters, which render the model more complex than other existing models. Nevertheless, the existence of four free parameters, make the model easily adjustable to the phenomenological and theoretical constraints. So, as we will see, early acceleration and late time acceleration are very conveniently described. In addition, local gravity constraints are satisfied and in addition a matter dominated epoch exists. Moreover, the stability of cosmological perturbations is ensured. The scalaron mass is positive and large and in addition, in the Einstein frame, the corresponding $\sigma$ field has a large and positive mass, thus rendering the Newton's law corrections negligible. In addition, the matter era constraints are satisfied, since the second derivative of the model is exponentially small. Finally, as we will see, the matter stability constraint, which is expressed in terms of an effective potential, is satisfied, since the effective potential is negative for a large range of values of the parameters. A fine tuning of the parameters is necessary, in order the aforementioned constraints are simultaneously satisfied. We have to add the fact that the model is just another viable exponential model and complements the existing models, but is somewhat more complex than some existing models, like the one described for example in reference \cite{paper1}.

\noindent This article is organized as follows: In section 1 we
briefly present the general features of $F(R)$ theories in the
Jordan frame within the context of the metric formalism. In
section 2 we study in detail the exponential model and also we
investigate all the criteria which have to be satisfied in order a
modified gravity model can be considered viable. In section 3 we
briefly present an interesting functional resemblance of two
$F(R)$ models (one of which is the one presented in this article),
to the fermi distributions connected to Woods-Saxons potentials.
The conclusions follow in the end of the article.

\section{General Features of $F(R)$ Dark Energy Models in the Jordan Frame}

In this section we shall give a brief description of the main
features of $F(R)$ theories in the Jordan frame. For a much more
detailed account consult references
\cite{reviews,importantpapers}. The $F(R)$ modified gravity
theories are described by the following four dimensional action:
\begin{equation}\label{action}
\mathcal{S}=\frac{1}{2\kappa^2}\int
\mathrm{d}^4x\sqrt{-g}F(R)+S_m(g_{\mu \nu},\Psi_m),
\end{equation}
with $\kappa^2=8\pi G$ and $S_m$ the matter action of the matter
fields $\Psi_m$. We shall focus on the metric formalism for our
study and in addition we shall assume that the form of the $F(R)$
theory that we shall present is of the form $F(R)=R+f(R)$. Within
the context of the metric formalism, by varying the action
(\ref{action}) with respect to $g_{\mu \nu}$, we obtain the
following equations of motion:
\begin{equation}\label{eqnmotion}
F'(R)R_{\mu \nu}(g)-\frac{1}{2}F(R)g_{\mu
\nu}-\nabla_{\mu}\nabla_{\nu}F'(R)+g_{\mu \nu}\square
F'(R)=\kappa^2T_{\mu \nu}.
\end{equation}
In the above equation, $F'(R)=\partial F(R)/\partial R$ and
$T_{\mu \nu}$ is the energy momentum tensor. The main idea behind
the $F(R)$ modified gravity theories is that, what is actually
modified is not the left hand side of the Einstein Equations, but
the right. Indeed, the above equations of motion can be cast in
the following form:
\begin{align}\label{modifiedeinsteineqns}
R_{\mu \nu}-\frac{1}{2}Rg_{\mu
\nu}=\frac{\kappa^2}{F'(R)}\Big{(}T_{\mu
\nu}+\frac{1}{\kappa}\Big{[}\frac{F(R)-RF'(R)}{2}g_{\mu
\nu}+\nabla_{\mu}\nabla_{\nu}F'(R)-g_{\mu \nu}\square
F'(R)\Big{]}\Big{)}.
\end{align}
Thus, the energy momentum tensor has another contribution coming
from the term:
\begin{equation}\label{newenrgymom}
T^{eff}_{\mu
\nu}=\frac{1}{\kappa}\Big{[}\frac{F(R)-RF'(R)}{2}g_{\mu
\nu}+\nabla_{\mu}\nabla_{\nu}F'(R)-g_{\mu \nu}\square
F'(R)\Big{]},
\end{equation}
and this term is what actually models the dark energy. Taking the
trace of equation (\ref{eqnmotion}) we obtain the following
equation:
\begin{equation}\label{traceeqn}
3\square F'(R)+R F'(R)-2F(R)=\kappa^2 T,
\end{equation}
with $T$ being the trace of the energy momentum tensor $T=g^{\mu
\nu}T_{\mu \nu}=-\rho+3P$, and $\rho$, $P$ are the energy density
and pressure of the matter respectively. This equation actually
shows us that there is another degree of freedom underlying the
$F(R)$ theories, materialized by the scalar field $F'(R)$.
Consequently, equation (\ref{traceeqn}) is actually the equation
of motion of this scalar degree of freedom which is called the
``scalaron''. In a flat Friedmann-Lemaitre-Robertson-Walker
spacetime, the Ricci scalar is given by:
\begin{equation}\label{ricciscal}
R=6(2H^2+\dot{H}),
\end{equation}
with $H$ the Hubble parameter, and the ``dot'' represents
differentiation with respect to time. Accordingly, the
cosmological dynamics are governed by the following equations:
\begin{align}\label{flrw}
&
3F'(R)H^2=\kappa^2(\rho_m+\rho_r)+\frac{(F'(R)R-F(R))}{2}-3H\dot{F}'(R),
\\ \notag &
-2F'(R)\dot{H}=\kappa^2(p_m+4/3\rho_r)+F\ddot{F}'(R)-H\dot{F}'(R),
\end{align}
where $\rho_r$ and $\rho_m$ stand for the radiation and matter
energy density respectively. Hence, the total effective energy
density and pressure of matter and geometry are
\cite{reviews,paper2}:
\begin{align}\label{densitypressure}
&
\rho_{eff}=\frac{1}{F'(R)}\Big{[}\rho+\frac{1}{\kappa^2}\Big{(}F'(R)R-F(R)-6H\dot{F}'(R)\Big{)}\Big{]}
\\ \notag &
p_{eff}=\frac{1}{F'(R)}\Big{[}p+\frac{1}{\kappa^2}\Big{(}-F'(R)R+F(R)+4H\dot{F}'(R)+2\ddot{F}'(R)\Big{)}\Big{]},
\end{align}
with $\rho,P$ the total matter energy density and pressure
respectively.

\section{General Study and Detailed Study of the Exponential Model}

In principle, every viable $F(R)$ model in the metric formalism
has to satisfy certain conditions, in order to be compatible with
observations and also consistent with theoretical predictions. In
this section we shall present a viable exponential model, which
has as limiting cases other exponential models that exist in the
literature, and we shall thoroughly study the quantitative
features of the $F(R)$ theory it describes. The exponential model
is of the following form,
\begin{equation}\label{mainmodel}
F(R)=R-\frac{C}{A+Be^{-R/D}}+\frac{C}{A+B},
\end{equation}
with $A,B,C,D$ constants. In the rest of this paper, we shall
denote $f(R)$, the second term on the right hand side of Eq.
(\ref{mainmodel}), that is:
\begin{equation}\label{eqnms1}
f(R)=-\frac{C}{A+Be^{-R/D}}+\frac{C}{A+B}.
\end{equation}

\subsection{General Features of the Model}

Let the Ricci scalar of the universe at the present epoch be
denoted as $R_0\simeq 10^{-66}\mathrm{eV}^2$ and additionally
denote as $\Lambda,\Lambda_I$ the cosmological constant of the
universe today and during the inflation period of the universe,
respectively. The general features of a viable $F(R)$ model are
the following \cite{reviews}:
\begin{itemize}
 \item $(i)$ $F'(R)>0$ for $R\geq R_0$. Moreover if $R_1$ is a de Sitter final attractor of the system, then this inequality has to hold true for $R\geq R_1$.

\item $(ii)$ $F''(R)>0$ for $R\geq R_0$. This restriction is
required for consistency with local gravity tests, for the
presence of a matter dominated epoch and for the stability of
cosmological perturbations.

\item $(iii)$ $F(R)=R-\Lambda_I$ for $R\rightarrow \infty$. This
is required in order inflation occurs in the universe.

\item $(i\nu)$ $F(R)=R-\Lambda$ for $R\rightarrow R_0$ and
$f(R_0)=-\Lambda$. This is required in order the late time
acceleration occurs.

\item $(\nu)$ $F(0)=0$, in order a flat spacetime solution exists.

\end{itemize}
In order model (\ref{mainmodel}) satisfies the above constraints,
the constants have to satisfy some conditions. In order to have a
clear picture of how the derivatives behave as a function of $R$,
we quote them below:
\begin{align}\label{firstsecondder}
&
F'(R)=1-\frac{BCe^{\frac{R}{D}}}{D\Big{(}B+Ae^{\frac{R}{D}}\Big{)}^2},\\
\notag &
F''(R)=\frac{BCe^{\frac{R}{D}}\Big{(}-B+Ae^{\frac{R}{D}}\Big{)}}{D^2\Big{(}B+Ae^{\frac{R}{D}}\Big{)}^3}.
\end{align}
Due to the square in the denominator of both the derivatives
$F'(R)$ and $F''(R)$, it is not very difficult to choose the
constants $A,B,C,D$. Indeed, by looking both expressions, the
first and second derivative are always positive when $A>B$ and
$D>C$. In order not to fall into inconsistencies, we also require
that the inequality means that the related constants have at least
$\sim 10\%$ difference in their scale, which for example means
that $A$ is ten times larger than $B$. Further restrictions on the
constants shall be imposed later on in this section. Hence, the
conditions $(i)$ and $(ii)$ in the list above are satisfied for
all the values of the Ricci scalar $R$, an argument that is
further supported from the fact that the denominator behaves as
$\sim e^{2R/D}$, rendering the fraction smaller than one for all
$R$, in reference to the first derivative. The second derivative
is always positive due to the fact that the term
$-B+Ae^{\frac{R}{D}}$ is always positive for $A>B$ due to the
exponential term. Condition $(\nu)$ is automatically satisfied for
the model (\ref{mainmodel}).

\noindent When $R\rightarrow \infty$, the function $F(R)$ behaves
as:
\begin{equation}\label{limitinfty}
F(R)\sim R-\frac{CB}{A\Big{(}A+B\Big{)}}.
\end{equation}
It is obvious that $\Lambda_I$, the early time cosmological
constant that governs the inflationary expansion of the universe,
must be $\Lambda_I\sim \frac{CB}{A(A+B)}$. Consequently, since
$\Lambda_I\sim 10^{20-38}\mathrm{eV}^2$, the values of $A,B,C$
have to be chosen appropriately in order this requirement is
satisfied. A set of values that will prove to be appropriate, in
order the present and other constraints that will be imposed later
on in this section are satisfied, are the following:
\begin{equation}\label{times}
A=0.1,{\,}B=0.01,{\,}C=10^{21},{\,}D=10^{22}.
\end{equation}
Using the values (\ref{times}), the inflation period cosmological
constant is approximately $\Lambda_I \sim 9.09\times 10^{20}$.
After checking these very general features, we proceed to more
elaborate criteria for the viability of the present exponential
model.

\noindent As a final comment before we proceed, notice that when
$R\gg R_0$, the model (\ref{mainmodel}), is approximately equal
to:
\begin{equation}\label{overlapwithm}
F(R)\sim
R-\frac{C}{A}+\frac{BC}{A^2}e^{-\frac{R}{D}}+\frac{C}{A+B}
\end{equation}
Therefore, we can see that the model of \cite{paper1,paper2} and
the model (\ref{mainmodel}) coincide in the large curvature limit.
Hence, we expect that these models may give rise to similar
cosmological dynamics. We shall elaborate on this issue later on
in this section.

\subsection{Scalaron Mass and Effective Potential}

The approach we adopt in the present section is based on
references \cite{paper1,paper2}. Following \cite{paper1,paper2},
the scalaron dynamical degree of freedom is governed by the trace
equation of motion (\ref{traceeqn}). By making the substitution
$F'(R)=1+f'(R)=e^{-\chi}$, and performing a perturbation around a
constant scalar curvature solution $R_*$, so that $R=R_*+\delta
R$, the equation of motion of the scalaron field read:
\begin{equation}\label{scalaroneqnmotion}
\square \delta
\chi-\frac{1}{3}\Big{(}\frac{1+f'(R_*)}{f''(R_*)}-R_*\Big{)}\delta
\chi= -\frac{\kappa^2}{6(1+f'(R_*))}T,
\end{equation}
with $\delta \chi$ being related to $\delta R$ as follows:
\begin{equation}\label{dsrcrrel}
\delta R=-\frac{1+f'(R_*)}{f''(R_*)}\delta \chi.
\end{equation}
The constant scalar curvature solution $R_*$, is a solution to the
equation:
\begin{equation}\label{desittsol}
R_*+2f(R_*)-R_*f'(R_*)=0.
\end{equation}
The mass square of the scalaron plays an important role in
reference to local and planetary test of the modified gravity
theory. This is equal to:
\begin{equation}\label{msscalaron}
M^2=\frac{1}{3}\Big{(}\frac{1+f'(R_*)}{f''(R_*)}-R_*\Big{)}.
\end{equation}
Only a positive and large value of the mass square is required, in
order no tachyonic instability occurs and the corrections to the
Newton's law are small. Substituting the model (\ref{eqnms1}), in
Eq. (\ref{desittsol}), we have that $R_*$ is a de Sitter solution
of the system. The value $R_*=0$ corresponds to the Minkowski
spacetime, so we would like to see if Minkowski spacetime is
stable. Particularly, we shall examine if the scalaron theory
perturbed around the Minkowski solution $R_*=0$ is stable, or if
it has tachyonic or any other sort of instabilities. This test is
critical for flat spacetime local tests. For $R_*=0$ and using the
values (\ref{times}), the mass of the scalaron field for the model
(\ref{eqnms1}), is equal to:
\begin{equation}\label{masscalval}
M^2(0)\sim 4.552\times 10^{22}.
\end{equation}
The above value is positive and consequently there is no tachyonic
instability. In addition, since it is very large, the $\delta R$
perturbations at long ranges tend to zero and therefore, Newton's
law has no considerable corrections. Actually, the mass square is
positive for a large range of values of the parameters, provided
that $A>B$ and $D>C$. This can be easily seen by looking the
analytic expression of the mass, for the model at hand:
\begin{equation}\label{masscghyrtyalval}
M^2(0)=\frac{(A+B)D\Big{(}-BC+(A+B)^2D\Big{)}}{3(A-B)BC}.
\end{equation}

\noindent Before closing this issue, we shall search for other de
Sitter solutions of Eq. (\ref{desittsol}). We are especially
interested in a late time de Sitter point. Following the notation
of \cite{paper2}, let $G(R)=F(R)-RF'(R)$. Since $G'(0)<0$ (see Eq.
(\ref{firstsecondder})), the function becomes negative and
increases after $R=R_0$. For $R=\mathcal{O}(\Lambda) $, the
derivatives of the $F(R)$ function behave as:
\begin{equation}\label{frfucntionbehavior}
F'(R)\simeq 1,{\,}{\,}{\,}F''(R)\simeq 0.
\end{equation}
The computation of the late time de Sitter point is
straightforward, since Eq. (\ref{desittsol}), simplifies to the
following:
\begin{equation}\label{simplifieddesitter}
2F(R)=R.
\end{equation}
Searching a solution around $R=\Lambda$ and using the values
(\ref{times}) for the variables $A,B,C,D$, we easily get
numerically, that $R=\Lambda$ is actually a late time de Sitter
solution.

\subsection{Matter Instability Analysis of the Exponential Model}

We now turn our focus on the matter instability issue, which might
occur when the scalar curvature is large, in reference to the one
corresponding to the present epoch. The scalaron equation can be
written in the form \cite{paper1,paper2}:
\begin{equation}\label{drperturb}
\square
R+\frac{f'''(R)}{f''(R)}\nabla_{\rho}R\nabla^{\rho}R+\frac{(1+f'(R))R}{3f''(R)}-\frac{2(R+f(R))}{3f''(R)}=\frac{\kappa^2}{6f''(R)}T.
\end{equation}
We perturb the scalar curvature around an Einstein gravity
solution $R_e=-\frac{\kappa^2T}{2}>0$, so that $R=R_e+\delta R$,
and we obtain the equation:
\begin{equation}\label{neweqn}
(-\partial^2_t+U(R_e))\delta R+\mathrm{const}\simeq 0,
\end{equation}
with $U(R_e)$ the effective potential, which is:
\begin{align}\label{effectpotential}
 U(R_e)&=\Big{(}\frac{F''''(R_e)}{F''(R_e)}-\frac{F'''(R_e)^2}{F''(R_e)^2}\Big{)}\nabla_{\rho}R_e\nabla^{\rho}R_e+\frac{R_e}{3}\\ \notag &
-\frac{F'(R_e)F'''(R_e)R_e}{3F''(R_e)^2}-\frac{F'(R_e)}{3F''(R_e)^2}+\frac{2F(R_e)F'''(R_e)}{3F''(R_e)^2}-\frac{F'''(R_e)R_e}{3F''(R_e)^2}.
\end{align}
In the case the effective potential is positive, the perturbation
$\delta R$ becomes exponentially large and the whole system is
rendered unstable \cite{paper1,paper2}. Therefore, the matter
stability condition is $U(R_e)<0$. Let us see what happens in the
case of the exponential model at hand (\ref{mainmodel}). If we
substitute (\ref{mainmodel}) to Eq. (\ref{effectpotential}), and
keep only the leading order terms, the potential $U(R)$ can be
written as follows:
\begin{equation}\label{potentialformre}
U(R_e)\sim
\frac{D^2e^{-\frac{R_e}{D}}\Big{(}B+Ae^{\frac{R_e}{D}}\Big{)}^3}{3BC\Big{(}B-Ae^{\frac{R_e}{D}}\Big{)}}.
\end{equation}
Due to the term in the denominator, namely $B-Ae^{\frac{R}{D}}$,
and since we imposed the condition $A>B$ for the viability of the
model, the potential given by relation (\ref{potentialformre}) is
always negative, for all curvature values and for all the values
of the parameters so that, $A>B$ and $D>C$. Thereby, the matter
stability condition is fulfilled for the model (\ref{mainmodel}).

\noindent Before finishing this section we would to address an
issue related to the matter era period of the cosmological model.
Following \cite{paper2}, during the matter era period, and
neglecting the contribution of radiation, one has:
\begin{equation}\label{mattereracond}
\rho_{eff}=\frac{\rho}{F'(R)},{\,}{\,}{\,}\frac{RF'(R)}{F(R)}=1,
\end{equation}
with $\rho_{eff}$ and $\rho$ the total energy density (matter and
dark energy) and $\rho$ the matter energy density. From the second
relation of Eq. (\ref{mattereracond}) we get \cite{paper2}:
\begin{equation}\label{secondeqncond}
\frac{F''(R)}{F'(R)}=0.
\end{equation}
So at the matter era we must have that $F''(R)\simeq 0$. As
pointed in \cite{paper2}, an $F(R)$-theory is acceptable if the
modified gravity contribution vanishes during this era and also
$F'(R)\sim 1$. In addition, the second derivative must be very
small and positive \cite{paper2}. Let us examine what happens for
the model (\ref{mainmodel}). As can be seen, the second derivative
of the $F(R)$ exponential is governed by the exponential term in
the denominator, that is:
\begin{equation}\label{mattereracondition}
F''(R)\simeq \frac{BC}{D^2A^2e^{\frac{R}{D}}}.
\end{equation}
Using the values (\ref{times}) for the parameters $A,B,C,D$ and
due to the exponential dependence, it can be easily checked that
the second derivative during the matter era is very close to zero
and positive.

\subsection{Einstein Frame Analysis of the Model}

Apart from the latter approach we adopted in order to address the
problem of matter instability, there is another approach for this
problem which is related to the Einstein frame of the $F(R)$
theory. Hence, we shall investigate the problem of matter
instability and Newton's law corrections, in the Einstein frame.
The Jordan frame and the Einstein frame description of $F(R)$
modified theories of gravity are mathematically equivalent.
Actually, the $F(R)$ theories in the Jordan frame become scalar
tensor theories with a potential in the Einstein frame
\cite{reviews,singularitiesaccelerationmodels,paper1,paper2}.
Following reference \cite{paper1}, an auxiliary field
$\mathcal{A}$ is introduced in the action (\ref{action})
\begin{equation}\label{einsteinframeaction}
\mathcal{S}=\frac{1}{\kappa^2}\int
\mathrm{d}x^4\sqrt{-g}\Big{(}(R-\mathcal{A})(1+f'(\mathcal{A}))+\mathcal{A}+f(\mathcal{A})\Big{)}.
\end{equation}
Using the conformal transformation $g_{\mu \nu}\rightarrow
e^{\sigma}g_{\mu \nu}$, with $\sigma =-\ln(1+f'(\mathcal{A}))$,
and bearing in mind that the equations of motion with respect to
$\mathcal{A}$, yield the result that $\mathcal{A}=R$, we obtain
the Einstein frame action:
\begin{align}\label{einsteinframe}
\mathcal{S}_E=\frac{1}{\kappa^2}\int \mathrm{d}x^4\sqrt{-g}
\Big{(}R-\frac{3}{2}g^{\mu
\nu}\partial_{\mu}\sigma\partial_{\nu}\sigma -V(\sigma)\Big{)},
\end{align}
with the effective potential $V(\sigma )$, being equal to:
\begin{equation}\label{vsigmapot}
V(\sigma
)=e^{\sigma}g(e^{-\sigma})-e^{2\sigma}F(g(e^{-\sigma}))=\frac{\mathcal{A}}{F'(\mathcal{A})}-\frac{F(\mathcal{A})}{F'(\mathcal{A})^2}
\end{equation}
The mass corresponding to the field $\sigma$ is equal to:
\begin{equation}\label{msasquareeinstein}
m_{\sigma}^2=\frac{1}{2}\frac{\mathrm{d}^2V(\sigma)}{\mathrm{d}\sigma^2}=\frac{1}{2}\Big{(}\frac{\mathcal{A}}{F'(\mathcal{A})}-\frac{4F(\mathcal{A})}{F'(\mathcal{A})^2}+\frac{1}{F''(\mathcal{A})}\Big{)}.
\end{equation}
In order the Newton's law corrections are small, this mass term
has to be a large number and of course positive in order tachyonic
instabilities are avoided. Let us see how this term behaves.
Substituting Eq. (\ref{mainmodel}) into (\ref{msasquareeinstein}),
the mass of the $\sigma$ field is equal to (we keep only
dominating terms):
\begin{equation}\label{fmsquareeinstei}
m_{\sigma}^2\simeq \frac{D^2A^2e^{\frac{2R}{D}}}{2BC}
\end{equation}
Let us find the value of the above mass, by using the values
(\ref{times}) for the parameters. For a wide range of values of
the scalar curvature, the mass (\ref{fmsquareeinstei}) has the
value $m_{\sigma}^2\sim 7.2\times 10^{22}$, which is large, and
thus we conclude that the Newton's law corrections are negligible.
In addition, since it is positive, the theory is free from
tachyonic instabilities.

\subsection{Analysis of Finite Time Singularities of the Exponential Model}

In this section we shall analyze in detail the singularity
structure of the model at hand. The singularities that quite
frequently occur in $F(R)$ theories of gravity have the form
$H(t)=h/(t_0-t)^{\beta}$, with $h$ and $t_0$ positive constants
\cite{paper2}. There are four types of finite time future
singularities
\cite{reviews,singularitiesaccelerationmodels,paper2} which are
listed below:
\begin{itemize}

 \item Type I (Big Rip): For $t\rightarrow t_0$, $a(t)\rightarrow \infty $, $\rho_{eff}\rightarrow \infty$ and $ |p_{eff}| \rightarrow \infty $ ($\beta \geq 1$)

\item Type II (sudden): For $t\rightarrow t_0$, $a(t)\rightarrow
a_0 $, $\rho_{eff}\rightarrow \rho_0$ and $ |p_{eff}| \rightarrow
\infty $ ($-1< \beta <0$)

\item Type III: For $t\rightarrow t_0$, $a(t)\rightarrow a_0 $,
$\rho_{eff}\rightarrow \infty$ and $ |p_{eff}| \rightarrow \infty
$ ($0<\beta <1$)

\item Type IV: For $t\rightarrow t_0$, $a(t)\rightarrow a_0 $,
$\rho_{eff}\rightarrow 0$ and $ |p_{eff}| \rightarrow 0 $ and
higher derivatives of the Hubble parameter $H$ diverge.

\end{itemize}

The singularity structure of the $F(R)$ model (\ref{mainmodel}) is
quite similar to the one presented in reference \cite{paper2}.
Indeed, at the limit $R\rightarrow \infty$, we have:
\begin{equation}\label{limitinfty}
\lim_{R\rightarrow \infty}F(R)\simeq
R+\mathrm{const},{\,}{\,}{\,}\lim_{R\rightarrow \infty}F'(R)\simeq
1.
\end{equation}
In addition, the higher order derivatives of the $F(R)$ function
(\ref{mainmodel}), tend to zero exponentially. Therefore, neither
Type I nor Type III singularities can appear in the present model,
which is exactly what happens in reference \cite{paper2}. Now let
us consider the other two possible singularities, namely Type II
and Type IV. In reference to the first, when $R\rightarrow
-\infty$, the $F(R)$ function can be written as:
\begin{equation}\label{snatch}
F(R)\simeq R-\frac{C}{B}e^{-R/D}+\frac{C}{A+B}.
\end{equation}
Consequently, the total effective energy density $\rho_{eff}$ of
Eq. (\ref{densitypressure}), exponentially decays for the model
(\ref{mainmodel}) and also the total effective pressure behaves
analogously. Thereby, the Type II singularity cannot occur, since
the pressure does not diverge. In addition, Type IV singularities
cannot be realized. The reason is in absolute concordance with the
arguments of \cite{paper2}. Indeed, when $R\rightarrow 0$, the
$F(R)$ function behaves as
\begin{equation}\label{frzweor}
F(R)\simeq R-\frac{BC}{(A+B)^2D}R,
\end{equation}
and therefore, the effective energy density behaves as $\sim
1/(t_0-t)^{\beta+1}$ and is larger than $ 1/(t_0-t)^{2\beta}$, for
$\beta <-1$. Hence the Type IV singularity cannot occur. As in the
\cite{paper2} case, the model we presented is free of
singularities. The only difference between the two models is the
different reasoning for the Type II singularity, since in the
present model it is the finite pressure that renders the model
free of these singularities, in contrast to the model of
\cite{paper2}, where the effective energy density diverges. In
addition, since according to Eq. (\ref{overlapwithm}), the two
models overlap when $R\rightarrow \infty$, the reasoning on why
the Type I and Type II singularities do not occur in both models
is identical in both cases, as it was expected.

\section{Comparison With Other Models and a Brief Discussion}

The model we investigated in this paper is a dark energy model
that belongs to the $\Lambda$CDM class of models. However, this
model is more complicated from other existing exponential models,
owing to the fact that it has four free parameters and hence a
fine-tuning is necessary in order to be phenomenologically and
theoretically viable. Nevertheless, due to the fine tuning, it
offers good phenomenological results. In this section we shall
briefly present some of the exponential models appearing in
references \cite{paper1,paper2,exponentialmodels} and compare
these to our model.

\subsection{The Model $F(R)=R-2\Lambda (1-e^{-R/R_0})$}

One of the most successful exponential models is the one studied
in detail in \cite{paper2}, which is:
\begin{equation}\label{descr1}
F(R)=R-2\Lambda (1-e^{-R/R_0}).
\end{equation}
This model combines al the good characteristics of a viable
$\Lambda$CDM, which we now describe in brief. As we described in
the previous sections, in order a model is viable, a large and
positive value of the scalaron mass squared is required. If this
is so, then the Minkowski spacetime solution $R_*$ is stable. The
Minkowski solution is a de-Sitter solution of the model
(\ref{descr1}) and for specific values of the parameters
$(R_0,\Lambda )$, the mass square (\ref{msscalaron}) is positive
and large. For the same values the early cosmological constant and
the late time acceleration are described in a successful way.
Moreover, in the Einstein frame, the mass of the $\sigma $ field
is large and positive, and in effect, the corrections to Newton's
law are negligible. Moreover, the matter instability condition for
the potential (\ref{effectpotential}), is fulfilled when $R_0<4
\Lambda$, and the potential is:
\begin{equation}\label{descr2}
 U(R_e)=-\frac{R_0e^{R/R_0}(R_0-4\Lambda)+2R_0\Lambda}{6\Lambda}
\end{equation}
Finally, the second derivative of the $F(R)$ function is equal to:
\begin{equation}\label{descr3}
F''(R)=\frac{2\Lambda}{R_0^2e^{R/R_0}}
\end{equation}
and so it is very small and positive, due to the exponential
dependence. Consequently, the second derivative during the matter
era is very close to zero and positive. In conclusion, the model
(\ref{descr1}) passes all the theoretical and phenomenological
tests, as our model does, but it is described with only two free
variables, which makes it more promising than our model. In
addition, this model (\ref{descr1}), almost coincides with the one
we studied in this paper, in the large $R$ limit.

\subsection{Brief Discussion}

Our model has many similarities to the four parameter exponential
model that appears in \cite{paper1},
\begin{equation}\label{4parmod}
F(R)=R-a(e^{-bR-1}+cR^N\frac{e^{bR}-1}{e^{bR}-e^{bR_I}})
\end{equation}
with $a$, $b$, $N$, and $R_I$, free parameters. The model
(\ref{4parmod}) is refinement of the model
\begin{equation}\label{4pajrmod}
F(R)=R-a(e^{-bR-1})
\end{equation}
It appears that our model is complementary to the most popular
exponential models due to the existence of too many parameters.
The advantage that our model offers is a better control of the
complete alignment to observations, at expense of having too many
free variables describing the model. This explains the fact that
the values of $C$ and $D$ are so huge in reference to those of $A$
and $B$, that is, in order to satisfy theoretical and
phenomenological constraints. Nevertheless, what actually
motivated us to study this model, is that it originates from some
nuclear potentials and hence we wanted to draw attention to these
models and investigate if there are any other models that offer
good phenomenology. In the next section we describe from which
nuclear models was the present study motivated.

\section{Functional Resemblance of two $F(R)$ Models to Wood-Saxons Nuclear Potentials}

The model we used in this article described by the $F(R)$ function
(\ref{mainmodel}), has many similarities to some Woods-Saxons
potentials that are used to describe the Nucleon interactions. The
Woods-Saxon potentials are phenomenological potentials for the
nucleons in the atomic nucleus. It is used in the shell model and
describes the forces applied to each nucleon. The general form of
the potential is,
\begin{equation}\label{woodssaxon}
V(r)=-\frac{V_0}{1+e^{\frac{r-b}{\alpha}}},
\end{equation}
where $r$ the distance from the center of the nucleus. Note one
particular fermi function distribution, related to
(\ref{woodssaxon})
\begin{equation}\label{fermicol}
u_f(r)\simeq B_c\tanh\big{(}\frac{-r}{2\alpha}\big{)}.
\end{equation}
Notice that this function (\ref{fermicol}) has great similarities
to the Tsujikawa model \cite{importantpapers,reviews}, for $F(R)$
modified gravitational theories, namely:
\begin{equation}\label{}
F(R)=R-R_c\tanh (R/R_c).
\end{equation}
Indeed, if we change $r\rightarrow -R$, and add $R$ in expression
(\ref{fermicol}) we get the Tsujikawa model. Using the same line
of reasoning, the following Fermi shape distribution,
\begin{equation}\label{fermic}
u_f(r)=\frac{1}{A+Be^{\frac{r}{\alpha}}},
\end{equation}
leads to the model we used in this article. Of course there is no
physical connection between the two systems, the nucleus potential
and the gravitational systems. We wanted to stress the fact that
by performing the same transformations to two different nuclear
fermi distributions, we get two viable $F(R)$ dark energy models.

\section*{Concluding Remarks}

In this article, we studied in detail an exponential model of
$F(R)$ gravity in the metric formalism. Apart from the very
general features that make the model viable at a first step, the
model has also other interesting features that render it
promising. Particularly, the scalaron mass is positive and large,
which means that the theory is free of tachyonic instabilities and
free of matter instabilities. The matter instability and Newton's
law corrections problem has been addressed in both the Jordan and
Einstein frames, and the result verified once more that the model
does not lead to problems or inconsistencies. Moreover, we studied
the finite time singularities issue and as we demonstrated, the
model is free of the four types of finite time singularities. In
addition, we also presented what motivated us to use such an
exponential model. As we showed, this model and another $F(R)$
model have functional resemblance to two Fermi distributions
corresponding to Woods-Saxons potentials.

\noindent We intended to show the basic features of this $F(R)$
model, using a very general approach, similar to the one used in
the articles \cite{paper1,paper2}. It is interesting to address
other interesting issues, such as oscillations of the dark energy,
or curvature singularities in dense gravitational backgrounds
\cite{tel1}, which however are beyond the scope of this article.
We hope to address such issues in a future work.

\end{document}